\documentclass[aps,prd, preprint, showpacs]{revtex4-1}
\usepackage{amsmath}
\usepackage{indentfirst}
\usepackage{graphicx}
\usepackage[all]{xy}
\usepackage{multirow, amssymb, wasysym, gensymb}

\begin{document}

\title{The equivalent vector boson approximation at threshold energies}

 \author{I. Alikhanov}
  \email{ialspbu@gmail.com}
   \affiliation{Institute for Nuclear Research of the Russian Academy of Sciences, 117312 Moscow, Russia\\Institute for Applied Mathematics and Automation, 360000 Nalchik, Russia}

\begin{abstract} 
A simple derivation of the distributions of the equivalent electroweak vector bosons in leptons and quarks  is presented. The applicability of the equivalent vector boson approximation at relatively low energies close to the $W$ and $Z$ boson thresholds is demonstrated. It is shown that the threshold correction to the distribution functions emerges naturally in the theory. Implications of the results for processes with the emission of the gauge bosons are discussed. 
\end{abstract}

\maketitle

\section{Introduction}
The quark--parton model (QPM)~\cite{Bjorken:1969ja}  is an effective tool for description of deep inelastic lepton--hadron and hadron--hadron collisions and widely used in modern particle physics. The QPM represents  the observed cross section as a sum of lepton--quark, quark--quark, gluon--quark or gluon--gluon subprocess cross sections by introducing parton distribution functions which parametrize the probability densities of finding quarks or gluons in a hadron with a given fraction of the parent hadron momentum. Along with the quantitative success, the QPM provides a relatively simple and intuitively conceivable view of interactions in such a way that it becomes possible to study various static and dynamical properties of particles unavailable in free states (as quarks and gluons).

One can extend the notion of the parton distributions and introduce similar functions for the photon and charged leptons~\cite{Chen:1975sh}. It is well known that at some conditions an electrically charged particle can be replaced by a flux of equivalent photons interacting with the target (the Weizs\"acker--Williams approximation). For example, knowing the momentum distribution of the equivalent photons in the electron allows us to probe two photon processes at electron--positron colliders~\cite{Brodsky:1971ud,Walsh:1973mz,Budnev:1974de} though we do not have real photons in the initial state. The electron acts as an effective photon. Conversely, the photon is also able to manifest itself in interactions as a charged lepton or quark and this can be parametrized by the corresponding lepton (quark) densities in the photon~\cite{kessler,Baier:1973ms}. The latter property may have interspersing implications for studies of lepton--lepton processes because it provides an effective beam of unstable muons and tau leptons that are a challenge to accelerate in ordinary experiments~\cite{Alikhanov:2014uja,Alikhanov:2015kla}.

The discovery of the massive electroweak vector bosons~\cite{Arnison:1983rp,Banner:1983jy,Arnison:1983mk,Bagnaia:1983zx} stimulated developing an analog of the equivalent photon approximation for $W$ and $Z$~\cite{Cahn:1983ip,Kane:1984bb,Dawson:1984gx}. Treating these bosons as equivalent particles can simplify calculations for processes probing the non-abelian gauge symmetry and the Higgs sector of the Standard Model through $WW$, $ZZ$ and $WZ$ scattering at $e^+e^-$ and $pp$ colliders~\cite{Boos:1997gw}. A certain amount of investigation has been devoted to the validity of the equivalent electroweak vector boson approximation (EVBA)~\cite{Kleiss:1986xp, Kunszt:1987tk,Accomando:2006mc,Alboteanu:2008my,Borel:2012by}, its improvements~\cite{Lindfors:1985yp, Johnson:1987tj, Kuss:1995yv,Kuss:1996ww,Bernreuther:2015llj} and applications for calculating electroweak and QCD corrections to boson scattering~\cite{Accomando:2006hq,Bouayed:2007rt}. 

In this article we focus on a class of reactions with the on-shell bosons in the final state that incorporate the equivalent particle approximation. An analysis of these reactions allows us to define the equivalent boson distributions in leptons/quarks not only at high energies but also close to the boson production thresholds, $\sqrt{s}\gtrsim m_{W,Z}$. This also leads to noticeable conclusions about the production mechanisms of the final state particles.

The article is organized as follows. In Section~\ref{sec1} we remind the main properties of the equivalent photon and electron densities in charged particles and in the photon, respectively. We also give general expressions of the cross sections for resonance production processes accompanied by initial state radiation calculated within the EPA. In Section~\ref{sec2} we compare straightforward calculations of the cross sections with the EPA approach. Section~\ref{section_ZZ} is devoted to the derivation of the equivalent $W$ and $Z$ boson distributions. Section~\ref{concl} contains the conclusions.

\section{General discussion\label{sec1}}

In the equivalent photon approximation (EPA) the probability density of finding a photon inside an electron with fraction $x$ of the parent electron momentum can be written as~\cite{Frixione:1993yw}

\begin{equation}
f_{\gamma/e}(x,Q^2)=\frac{\alpha}{2\pi}\left[\frac{1+(1-x)^2}{x}\ln\left(\frac{Q^2_{\text{max}}}{Q^2_{\text{min}}}\right)+\mathcal{O}(1)\right],\label{eq:gen1}
\end{equation}
where $\alpha=e^2/(4\pi)$ is the fine structure constant, $Q^2_{\text{min}}$ and $Q^2_{\text{max}}$ are the minimum and maximum of the magnitude of the four-momentum transfer in the given process. The non-logarithmic term is, in general, also a function of $x$ and $Q^2$. On the other hand, in the massless limit the outgoing electron carries the fraction $y=1-x$ of its initial momentum so that the probability density of finding an electron in an electron after the emission of the photon becomes

\begin{equation}
f_{e/e}(y,Q^2)=\frac{\alpha}{2\pi}\left[\frac{1+y^2}{1-y}\ln\left(\frac{Q^2_{\text{max}}}{Q^2_{\text{min}}}\right)+\mathcal{O}(1)\right].\label{eq:gen2}
\end{equation}
Equations \eqref{eq:gen1} and \eqref{eq:gen2} turn into each other under the exchange $y\leftrightarrow 1-x$ as it should be due to the energy-momentum conservation. Since the photon is able to convert into an $e^+e^-$ pair, one can introduce a similar distribution for electrons/positrons inside the photon as well~\cite{kessler,Baier:1973ms}:

\begin{equation}
f_{e/\gamma}(y,Q^2)=\frac{\alpha}{2\pi}\left[\left[y^2+(1-y)^2\right]\ln\left(\frac{Q^2_{\text{max}}}{Q^2_{\text{min}}}\right)+\mathcal{O}(1)\right].
\end{equation}
A detailed discussion of these functions can be found in~\cite{Chen:1975sh}.

\begin{figure}
\includegraphics[width=1.1\textwidth]{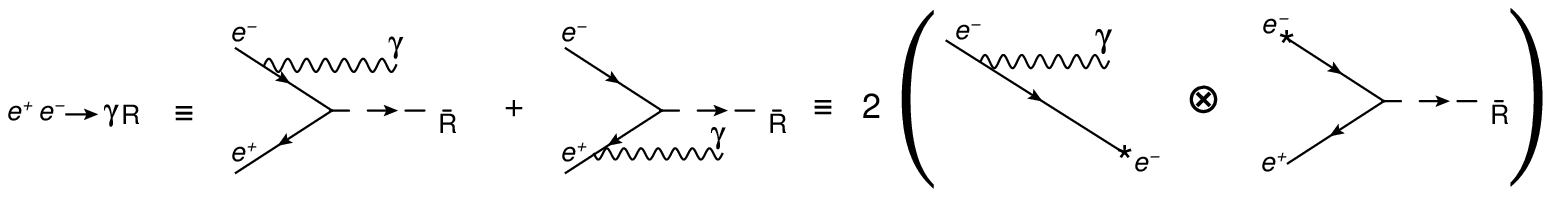}
\caption{The initial state radiation mechanism for the production of a resonance $R$ in $e^+e^-\rightarrow \gamma R$. The convolution of the equivalent electron density in the electron after the photon radiation with the cross section for the resonant subprocess $e^+e^-\rightarrow R$ is  symbolically represented by $\otimes$. The equivalent electron is tagged by an asterisk.}
\label{fig1p}
\end{figure}

\begin{figure}
\includegraphics[width=0.7\textwidth]{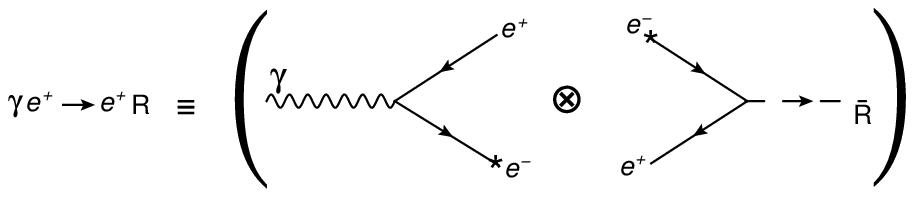}
\caption{The photon splitting mechanism for the production of a resonance $R$ in $\gamma e^+\rightarrow e^+R$. The convolution of the equivalent electron density in the photon with the cross section for the resonant subprocess $e^+e^-\rightarrow R$ is  symbolically represented by $\otimes$. The equivalent electron is tagged by an asterisk.}
\label{fig2p}
\end{figure}

As an example of the application of the EPA, let us consider the  production of an electrically neutral narrow resonance $R$ of mass $m_R$ in the $s$-channel electron--positron annihilation, $e^+e^-\rightarrow R$, at the center-of-mass (cms) energy $\sqrt{s}$. Assume that the production is accompanied by the initial state radiation, $e^+e^-\rightarrow \gamma R$ (see Fig.~\ref{fig1p}). Then the corresponding total cross section reads

\begin{equation}
\sigma_{\gamma R}(s)=2\int\limits_0^{1-\tau}f_{\gamma/e}(x,Q^2)\hat \sigma ((1-x)s)dx=2\int\limits_{\tau}^{1}f_{e/e}(y,Q^2)\hat \sigma (ys)dy, \label{eq:int1}
\end{equation}
where $\tau=m_R^2/s$, $\hat\sigma(s)$ is the cross section for the subprocess $e^+e^-\rightarrow R$. The factor 2 appears because the distributions for the electrons and positrons are equal. If the narrow width approximation is applicable to the resonance, then the subprocess cross section can be replaced by the delta function $\hat \sigma (ys)=a\delta(y-\tau)/s$ and~\eqref{eq:int1} leads to
%
%
\begin{equation}
\sigma_{\gamma R}(s)=\frac{2a}{s}f_{e/e}(\tau,Q^2). \label{eq:int2}
\end{equation}
Explicitly $a=4\pi^2(2J+1)\Gamma_{ee}/m_R$ with $J$ and $\Gamma_{ee}$ being the spin of the resonance and the $R\rightarrow e^+e^-$ decay width, respectively~\cite{okun}.
There is another possibility of the resonance excitation, in $\gamma e^+\rightarrow e^+R$, when the photon serves as a source of electrons through $\gamma\rightarrow e^+e^-$ splitting as shown in Fig.~\ref{fig2p}. The cross section for this case is  

\begin{equation}
\sigma_{e R}(s)=\frac{a}{s}f_{e/\gamma}(\tau,Q^2). \label{eq:int3}
\end{equation}
For $\gamma e^-\rightarrow e^-R$ one obviously will have the same formula. It should be noted that the photon splitting mechanism is initiated not only by the incident electrically charged leptons but by neutrinos as well~\cite{Alikhanov:2008eu,Alikhanov:2008ca}.  

From~\eqref{eq:int2} and~\eqref{eq:int3} we see that a narrow resonance projects out the given distribution function in the total cross section~\cite{Chen:1975sh}. In other words, one can expect that measuring or calculating the cross section of such a process one simultaneously determines the distribution function up to a known constant factor and vice versa.

\section{Inelastic Compton scattering \label{sec2}}
Considering the mechanisms of the resonance production discussed above one may ask about the completeness of such an approach. There may be additional diagrams of the same order that should be also taken into account. In this section we present a class of reactions which intrinsically contain the EPA. Straightforward calculations of the cross sections for these  reactions using the standard electroweak Lagrangian are reproduced by the EPA with a remarkable precision. 

\begin{figure}
\includegraphics[width=.9\textwidth]{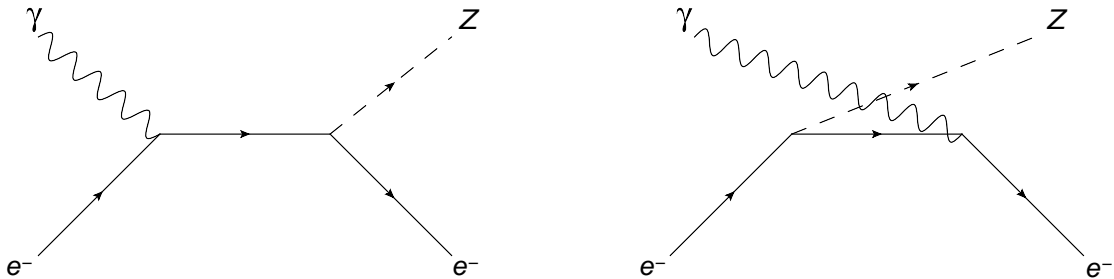}
\caption{The lowest-order Feynman diagrams that contribute to the inelastic Compton scattering $\gamma e^{-}\rightarrow e^{-} Z$.}
\label{fig3p}
\end{figure}

The first reaction is the inelastic Compton scattering $\gamma e^-\rightarrow e^- f$~\cite{Chen:1974wv}, where $f$ denotes a final state of non-zero mass. To be specific we take the massive neutral boson of the electroweak theory as~$f$: 

\begin{equation}
\gamma e^-\rightarrow e^- Z. \label{eq:int4}
\end{equation}
The lowest-order Feynman diagrams contributing to this process are shown in Fig.~\ref{fig3p}. In the limit of vanishing electron mass the Feynman rules with the standard electron--boson vertex $e\bar u_e\gamma^\mu(g_V-g_A\gamma^5)u_e/\sin2\theta_W$ yield the following square of the amplitude averaged over initial and summed over final spins:

\begin{equation}
|M|^2_{eZ}(s,t,u)=-\frac{32 \pi^2\alpha^2}{\sin^2{2\theta_W}} (g_V^2+g_A^2) \left[\frac{u}{s}+\frac{s}{u}+\frac{2 m_Z^2t}{su}\right],\label{amplv}
\end{equation}
where $\theta_W$ is the weak mixing angle, $g_V=-1/2+2\sin^2\theta_W$ and $g_A=-1/2$ are the vector and axial-vector couplings of the $Z$ boson to the electron, $t=-Q^2$ and $u=m_Z^2-s-t$ are the Mandelstam variables. Hence the total cross section for~\eqref{eq:int4} is
\begin{equation}
\sigma_{e Z}(s)=6\pi\frac{\alpha\Gamma_{ee}}{m_Zs}\left[\left[\tau^2+(1-\tau)^2\right]\ln\left(\frac{Q^2_{\text{max}}}{Q^2_{\text{min}}}\right)+h(Q^2)\right].\label{secm01}
\end{equation}
Here $\tau=m_Z^2/s$, $h(Q^2)=\left(Q^2_{\text{max}}-Q^2_{\text{min}}\right)\left(4m_Z^2+Q^2_{\text{max}}+Q^2_{\text{min}}\right)/(2s^2)$  and we have used the $Z\rightarrow e^+e^-$ decay width $\Gamma_{ee}=\alpha(g_V^2+g_A^2)m_Z/(3\sin^2{2\theta_W})$. Compare~\eqref{secm01} with the result of~\cite{Flambaum:1974wp}. As $Q^2_{\text{max}}\leq s-m_Z^2$, the non-logarithmic part in the square brackets is bounded: $0\leq h(Q^2)\leq1/2$.   It turns out that the cross section~\eqref{secm01} coincides with~\eqref{eq:int3} and we therefore conclude that the reaction~$\gamma e^-\rightarrow e^- Z$ proceeds through the resonant annihilation of the equivalent positrons from the incident photon splitting. A similar conclusion applies, of course, to the $CP$ conjugate process~$\gamma e^+\rightarrow e^+Z$.

\begin{figure}
\includegraphics[width=.9\textwidth]{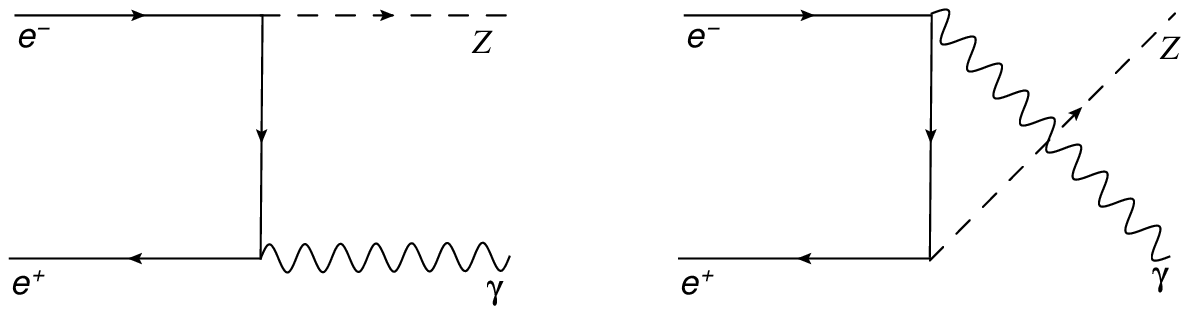}
\caption{The lowest-order Feynman diagrams that contribute to $e^+e^-\rightarrow \gamma Z$.}
\label{fig4p}
\end{figure}

The next example is bremsstrahlung in $e^+e^-\rightarrow \gamma f$~\cite{Chen:1974wv}. Again we consider the particular case $f=Z$: 

\begin{equation}
e^+e^-\rightarrow \gamma Z, \label{eq:int5}
\end{equation}
which is related to $\gamma e^-\rightarrow e^- Z$ by crossing symmetry. The relevant Feynman diagrams are shown in Fig.~\ref{fig4p}. One can simply derive the amplitude squared for~\eqref{eq:int5} by exchanging $s\leftrightarrow t$ in~\eqref{amplv}:
\begin{equation}
|M|^2_{\gamma Z}(s,t,u)=\frac{32 \pi^2\alpha^2}{\sin^2{2\theta_W}} (g_V^2+g_A^2) \left[\frac{u}{t}+\frac{t}{u}+\frac{2 m_Z^2s}{tu}\right].\label{amplv2}
\end{equation}
The overall factor $-1$ disappears because one fermion is crossed. See also~\cite{Berends:1986yy,Mery:1987et}. The corresponding cross section reads

\begin{equation}
\sigma_{\gamma Z}(s)=12\pi\frac{\alpha\Gamma_{ee}}{m_Zs}\left[\frac{1+\tau^2}{1-\tau}\ln\left(\frac{Q^2_{\text{max}}}{Q^2_{\text{min}}}\right)-\frac{1}{s}\left(Q^2_{\text{max}}-Q^2_{\text{min}}\right)\right].\label{secm02}
\end{equation}
Analyzing~\eqref{secm02} we notice an excellent agreement with~\eqref{eq:int2}. This allows us to interpret the reaction $e^+e^-\rightarrow \gamma Z$ as a process proceeding through the resonant annihilation of the equivalent electrons/positrons into the $Z$ boson as well. 

\begin{figure}
\includegraphics[width=.9\textwidth]{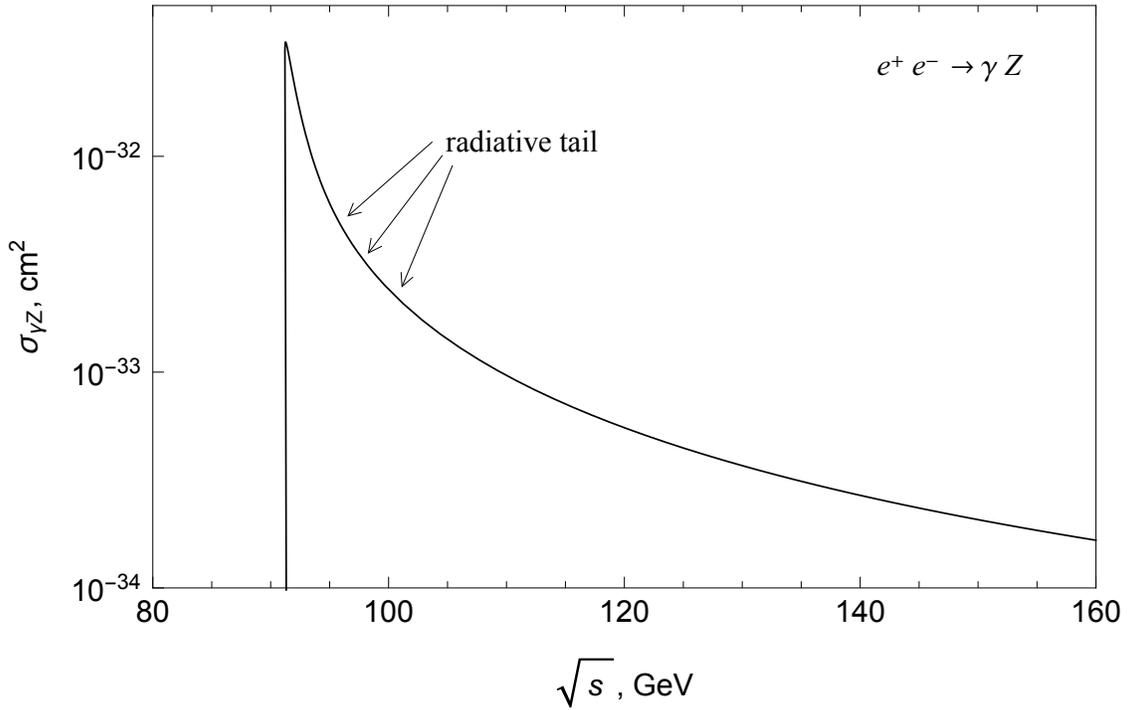}
\caption{The total cross section for $e^+e^-\rightarrow \gamma Z$ as a function of the center-of-mass energy. The radiative tail due to initial state $\gamma$ radiation is indicated by arrows.}
\label{fig5p}
\end{figure}

A visual analysis of the cross section as a function of the cms energy provides a relevant complement to our studies. Figure~\ref{fig5p} shows the radiative tail in the cross section which appears owing to initial state radiation. Even if the incident total energy of the electron--positron pair is above $m_Z$, the radiated photon carries away the "extra" energy and returns thus the collision energy to the resonance pole~\cite{Baier:1969kaa}. Formally  this is represented by the first integral in~\eqref{eq:int1}. This mechanism stretches the tail to the right of the resonance position clearly observed in $e^+e^-\rightarrow \gamma Z$. 

It should be emphasized that the presence of both diagrams in the amplitudes are required to ensure the reproduction of the EPA mechanisms in the cross sections. Had we neglected any of the diagrams, the resulting cross sections would have been significantly deviating form~\eqref{secm01} and~\eqref{secm02}.

\section{Massive boson radiation \label{section_ZZ}}
The discussion above has shown that the exact cross sections for the inelastic Compton scattering and bremsstrahlung when there is a resonance in the final state are in agreement with simple EPA calculations. Already the narrow width approximation gives a noticeable precision. 
It is interesting to analyze a process similar to $e^+e^-\rightarrow\gamma Z$ but with an additional massive boson instead of the final photon:

\begin{equation}
e^+e^-\rightarrow Z Z. \label{eezz}
\end{equation}
Being inspired by the application of the EPA to $e^+e^-\rightarrow\gamma Z$ we may hope that the EVBA will be successful for~\eqref{eezz} as well.  

\begin{figure}
\includegraphics[width=.9\textwidth]{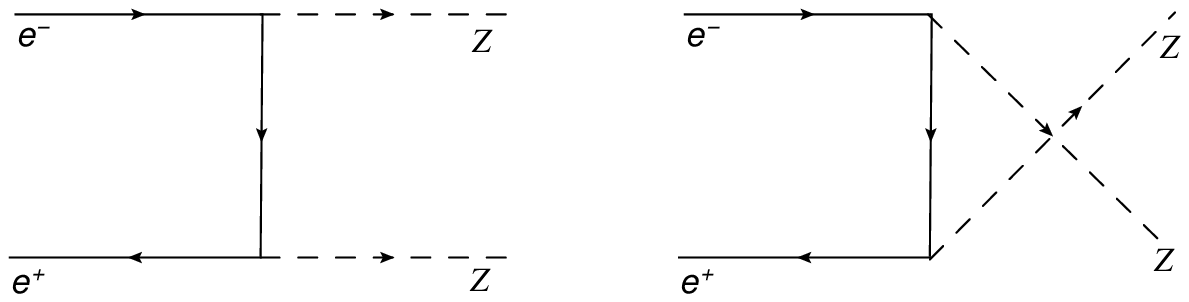}
\caption{The lowest-order Feynman diagrams that contribute to $e^+e^-\rightarrow ZZ$.}
\label{fig6p}
\end{figure}

\begin{figure}
\includegraphics[width=.9\textwidth]{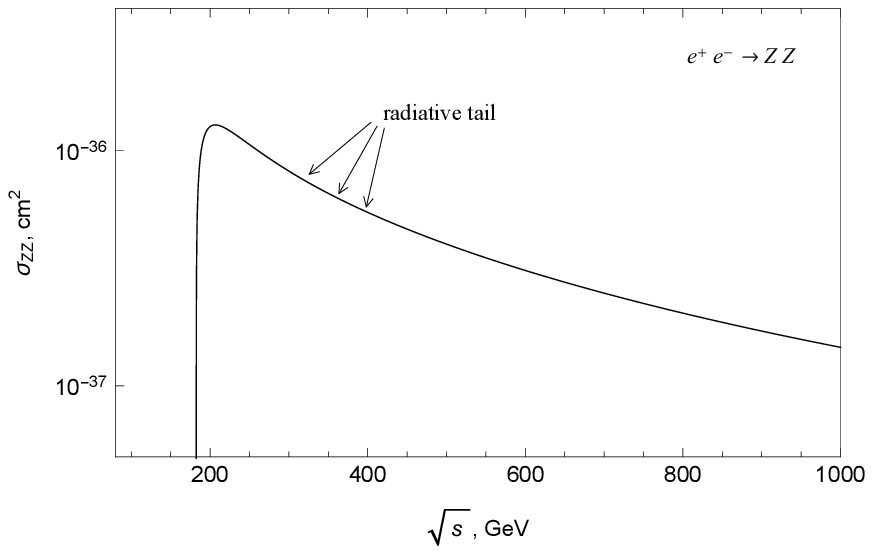}
\caption{The total cross section for $e^+e^-\rightarrow Z Z$ as a function of the center-of-mass energy. The radiative tail due to initial state $Z$ boson radiation is indicated by arrows.}
\label{fig7p}
\end{figure}

\begin{figure}
\includegraphics[width=1.1\textwidth]{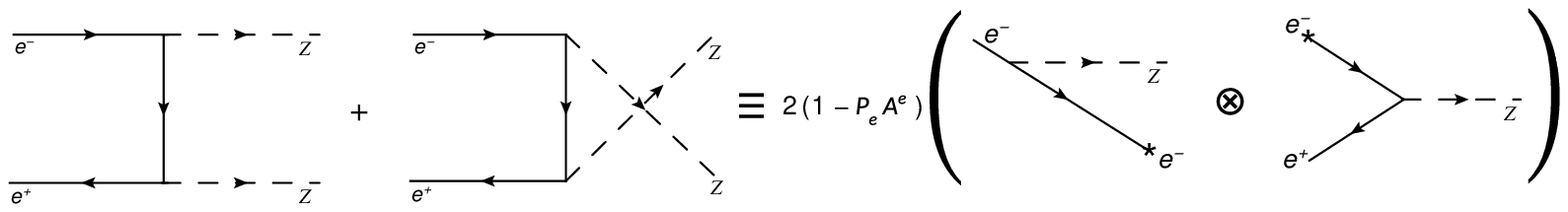}
\caption{Straightforward calculations of the cross section for $e^+e^-\rightarrow ZZ$  (the Feynman diagrams on the left hand side) give the same result as the initial sate $Z$ boson radiation mechanism (the right-hand side). The sign $\otimes$ represents the convolution of the equivalent electron density in the electron after the $Z$ boson radiation with the cross section for the resonant subprocess $e^+e^-\rightarrow Z$. The equivalent electron is tagged by an asterisk, $\mathcal{P}_e$ and $\mathcal{A}^e$ are the polarization of the equivalent electron and the asymmetry, respectively.}
\label{fig8p}
\end{figure}

As before, we begin with  the straightforward calculations in the framework of the electroweak theory. The corresponding Feynman diagrams depicted in Fig.~\ref{fig6p} lead to the following amplitude squared~\cite{Brown:1978mq}:

\begin{equation}
|M|^2_{Z Z}(s,t,u)=\frac{32 \pi^2\alpha^2}{\sin^4{2\theta_W}} \left(g_V^4+g_A^4+6g_V^2g_A^2\right) \left[\frac{u}{t}+\frac{t}{u}+\frac{4 m_Z^2s}{tu}-m_Z^4\left(\frac{1}{t^2}+\frac{1}{u^2}\right)\right]\label{amplzz}
\end{equation}
so that the total cross section can be presented in the form
\begin{equation}
\sigma_{ZZ}(s)=12\pi\frac{\alpha\Gamma_{ee}}{m_Zs}\frac{\left(g^2_V+g^2_A\right)}{\sin^2{2\theta_W}}\left(1+\frac{4g_V^2g_A^2}{(g_V^2+g_A^2)^2}\right)\left[\frac{1+4\tau^2}{1-2\tau}\ln\left(\frac{Q^2_{\text{max}}}{Q^2_{\text{min}}}\right)-\frac{1}{s}\left(Q^2_{\text{max}}-Q^2_{\text{min}}\right)\right].
\label{secm}
\end{equation}
An indirect hint in favor of the assumption that the initial state radiation mechanism takes place in $e^+e^-\rightarrow Z Z$ is a feature of its cross section resembling a radiative tail (see Fig.~\ref{fig7p}). 
If~\eqref{secm} is the projection of the distribution function of the equivalent electrons in the electron after the $Z$ boson radiation, then, in the limit of vanishing boson mass, $\sigma_{ZZ}(s)$ should be reduced to $\sigma_{\gamma Z}(s)$. However the possibility of the reduction is not obvious from~\eqref{secm} and requires some explanation. Analyze the logarithmic term. In the numerator $\tau^2$ is multiplied by 4 while in the denominator we have factor $-2$ in front of $\tau$. These factors have the same origin: now the energy of the equivalent electron is bounded from above in the sense that the latter cannot carry all the energy of the parent electron. The emitted $Z$ boson is massive and will always take away the fraction $\geq m_Z^2/s$ of the incident energy so that instead of $y$  we get $y+m_Z^2/s$ everywhere in the distribution function. Explicitly this threshold effect emerges in~\eqref{secm} through

\begin{equation}
\frac{1+4\tau^2}{1-2\tau}=\frac{1+(\tau+m_Z^2/s)^2}{1-(\tau+m_Z^2/s)}=\int\limits_{\tau}^1\frac{1+(y+m^2_Z/s)^2}{1-(y+m^2_Z/s)}\delta(y-\tau)dy.\label{th_effect}
\end{equation}
Recall that in the narrow width approximation the cross section for the resonant subprocess $e^+e^-\rightarrow Z$ is proportional to the delta function $\delta(y-\tau)$.
Now one can set the highlighted $Z$ boson mass in the logarithmic term equal to zero and thus reduce~\eqref{secm} exactly to~\eqref{secm02}. Let us also explain the physical meaning of the factor $1+4g_V^2g_A^2/(g_V^2+g_A^2)^2$ in~\eqref{secm}. This emerges because the equivalent electrons (positrons) get polarized in the $Z$ boson radiation process. In contrast to the photon, the $Z$ boson couples  with different strengths to the left and right-handed electrons that leads to the polarization $\mathcal{P}_e=-2g_Vg_A/(g_V^2+g_A^2)$ of the outgoing equivalent electrons. The cross section for electron--positron annihilation at the $Z$-pole, when the incident electron beam has polarization $\mathcal{P}_e$, reads $\sigma=(1-\mathcal{P}_e\mathcal{A}^e)\sigma_0$~\cite{Renton:1990td}. Here $\sigma_0$ is the unpolarized cross section,  $\mathcal{A}^e=2g_Vg_A/(g_V^2+g_A^2)$ is the asymmetry. Therefore, in the presence of the polarization,~\eqref{eq:int2} becomes

\begin{equation}
\sigma(s)=\frac{2a}{s}(1-\mathcal{P}_e\mathcal{A}^e)f_{e/e}(\tau,Q^2). \label{eq:int2f}
\end{equation}
Relying on~\eqref{th_effect} and comparing~\eqref{secm} with~\eqref{eq:int2f} we extract the distribution function for the equivalent electrons from~\eqref{secm}

\begin{equation}
f^T_{e/e}(y,Q^2)=\frac{\alpha}{2 \pi}\frac{\left(g^2_V+g^2_A\right)}{\sin^2{2\theta_W}}\left[\frac{1+\left(y+m_Z^2/s\right)^2}{1-y-m_Z^2/s}\ln\left(\frac{Q^2_{\text{max}}}{Q^2_{\text{min}}}\right)-\frac{1}{s}\left(Q^2_{\text{max}}-Q^2_{\text{min}}\right)\right],
\label{funfel}
\end{equation}
where $0\leq y\leq 1-m^2_Z/s$. This is graphically represented in Fig.~\ref{fig8p}. Note that only the equivalent transverse $Z$ bosons come into play in the presented example. From~\eqref{funfel}, it follows immediately that their distribution in the electron is
\begin{equation}
f^T_{Z/e}(x,Q^2)=\frac{\alpha}{2 \pi}\frac{\left(g^2_V+g^2_A\right)}{\sin^2{2\theta_W}}\left[\frac{1+\left(1-x+m_Z^2/s\right)^2}{x-m_Z^2/s}\ln\left(\frac{Q^2_{\text{max}}}{Q^2_{\text{min}}}\right)-\frac{1}{s}\left(Q^2_{\text{max}}-Q^2_{\text{min}}\right)\right],
\label{funfgam}
\end{equation}
where $m_Z^2/s\leq x\leq 1$.
Asymptotically  ($s\rightarrow\infty$)  this agrees with the original result of~\cite{Kane:1984bb,Dawson:1984gx} and coincides with~\eqref{eq:gen1} in the massless limit at $g_V^2/\sin^2{2\theta_W}=1$, $g_A^2=0$ as it should be. The $Z$ boson distributions inside the other leptons or quarks are also given by~\eqref{funfgam} provided the quark mass $m_q\ll m_{Z}$. In general, one should take $g_V=T^3_{f}-2Q_f\sin^2\theta_W$ and $g_A=T^3_{f}$, where standardly $T^3_f$ and $Q_f$ are the third component of the weak isospin and the electric charge of fermion $f$, respectively. Obviously, the distribution of the equivalent transverse $W^\pm$ bosons in the fermion will  have the same form:

\begin{equation}
f^T_{W/f}(x,Q^2)=\frac{\alpha}{8\pi\sin^2{\theta_W}}\left[\frac{1+\left(1-x+m_W^2/s\right)^2}{x-m_W^2/s}\ln\left(\frac{Q^2_{\text{max}}}{Q^2_{\text{min}}}\right)-\frac{1}{s}\left(Q^2_{\text{max}}-Q^2_{\text{min}}\right)\right],
\label{funfgam2}
\end{equation}
where we have used that for $W^{\pm}$, $g_V=g_A=e/(2\sqrt{2}\sin\theta_W)$. 

The necessity of the threshold correction to the distribution function of the $Z$ bosons as an extension of~\eqref{eq:gen1} is intuitively comprehensible and the ratio $m_{Z}^2/s$ could be inserted in~\eqref{eq:gen1} "by hands"~\cite{Alikhanov:2018jey} but appears naturally in the theory. It is therefore reasonable to assume that the same is also valid for the distributions of the equivalent longitudinal $Z$ and $W^\pm$. Since the asymptotic behavior of the distribution of the longitudinally polarized $Z$ bosons is~\cite{Kane:1984bb,Dawson:1984gx} 

\begin{equation}
\lim_{s\to\infty}f^L_{Z/f}(x)=\frac{\alpha}{\pi}\frac{\left(g^2_V+g^2_A\right)}{\sin^2{2\theta_W}}\frac{1-x}{x},
\label{funfl232}
\end{equation}
for all energies $\sqrt{s}\geq m_Z$ we obtain

\begin{equation}
f^L_{Z/f}(x)=\frac{\alpha}{\pi}\frac{\left(g^2_V+g^2_A\right)}{\sin^2{2\theta_W}}\frac{1-x+m_Z^2/s}{x-m_Z^2/s}
\label{funfl2}
\end{equation}
and, in analogy, for the longitudinal $W^{\pm}$ bosons

\begin{equation}
f^L_{W/f}(x)=\frac{\alpha}{4\pi\sin^2{\theta_W}}\frac{1-x+m_W^2/s}{x-m_W^2/s}.
\label{funfl}
\end{equation}

Thus, we have demonstrated that the reaction $e^+e^-\rightarrow ZZ$ can be interpreted as proceeding through initial state $Z$ boson radiation. This allows us to define the EVBA at relatively low energies comparable to the the masses of the bosons. The derived EVBA distribution functions have the poles $1/(x-m_{W,Z}^2/s)$ and predict that the leptons/quarks will preferably radiate equivalent bosons which are at rest ($x=m_{W,Z}^2/s$ means that the boson is at rest).

\section{Conclusions \label{concl}}
The equivalent massive vector boson approximation is usually applied to processes proceeding at such energies that allow one to neglect the boson masses ($\sqrt{s}\gg m_{W,Z}$). In this article we have shown that this approximation can be selfconsistently extended to relatively lower energies comparable to the masses of the bosons, $\sqrt{s}\gtrsim m_{W,Z}$. In the analysis we have used a class of reactions described by simple Feynman diagrams that intrinsically contain the equivalent particle approximation. We have derived the distributions of the equivalent $W$ and $Z$ bosons in leptons and quarks which take into account the energy thresholds of the boson production. The threshold correction to the distribution functions emerges naturally in the electroweak theory. In particular, our results lead to a noticeable conclusion about the mechanism of the reaction $e^+e^-\rightarrow ZZ$. The latter proceeds through the initial state $Z$ boson radiation in analogy with the well known initial state photon radiation in electron--positron annihilation into a resonance. Various implications of the initial state massive boson radiation for accelerator searches for physics beyond the Standard Model is discussed, e.g., in~\cite{Bednyakov:2015uoa}.

\begin{acknowledgments}
I thank E. A. Paschos for reading the manuscript. This work was partly supported by the Program of fundamental scientific research of the Presidium of the Russian Academy of Sciences "Physics of fundamental interactions and nuclear technologies".
\end{acknowledgments}



\newpage

\begin{thebibliography}{99}


\bibitem{Bjorken:1969ja} 
  J.~D.~Bjorken and E.~A.~Paschos,
  Phys.\ Rev.\  {\bf 185}, 1975 (1969).

\bibitem{Chen:1975sh} 
  M.~S.~Chen and P.~M.~Zerwas,
  Phys.\ Rev.\ D {\bf 12}, 187 (1975).

\bibitem{Brodsky:1971ud} 
  S.~J.~Brodsky, T.~Kinoshita and H.~Terazawa,
  Phys.\ Rev.\ D {\bf 4}, 1532 (1971).

\bibitem{Walsh:1973mz} 
  T.~F.~Walsh and P.~M.~Zerwas,
  Phys.\ Lett.\ B  {\bf 44}, 195 (1973).


\bibitem{Budnev:1974de} 
  V.~M.~Budnev, I.~F.~Ginzburg, G.~V.~Meledin and V.~G.~Serbo,
  Phys.\ Rept.\  {\bf 15}, 181 (1975).

\bibitem{kessler} 
P. Kessler, Nuovo Cim. {\bf 17}, 809 (1960).

\bibitem{Baier:1973ms} 
  V.~N.~Baier, V.~S.~Fadin and V.~A.~Khoze,
  Nucl.\ Phys.\ B {\bf 65}, 381 (1973).

\bibitem{Alikhanov:2014uja} 
  I.~Alikhanov,
  Phys.\ Lett.\ B {\bf 741}, 295 (2015)
  [arXiv:1402.6678 [hep-ph]].


\bibitem{Alikhanov:2015kla} 
  I.~Alikhanov,
  Phys.\ Lett.\ B {\bf 756}, 247 (2016)
  [arXiv:1503.08817 [hep-ph]].

\bibitem{Arnison:1983rp} 
  G.~Arnison {\it et al.} [UA1 Collaboration],
  Phys.\ Lett.\ B {\bf 122}, 103 (1983).

\bibitem{Banner:1983jy} 
  M.~Banner {\it et al.} [UA2 Collaboration],
  Phys.\ Lett.\ B {\bf 122}, 476 (1983).

\bibitem{Arnison:1983mk} 
  G.~Arnison {\it et al.} [UA1 Collaboration],
  Phys.\ Lett.\ B {\bf 126}, 398 (1983).


\bibitem{Bagnaia:1983zx} 
  P.~Bagnaia {\it et al.} [UA2 Collaboration],
  Phys.\ Lett.\ B {\bf 129}, 130 (1983).


\bibitem{Cahn:1983ip} 
  R.~N.~Cahn and S.~Dawson,
  Phys.\ Lett.\ B {\bf 136}, 196 (1984)
  Erratum: [Phys.\ Lett.\  B {\bf 138}, 464 (1984)].

\bibitem{Kane:1984bb} 
  G.~L.~Kane, W.~W.~Repko and W.~B.~Rolnick,
  Phys.\ Lett.\ B {\bf 148}, 367 (1984).

\bibitem{Dawson:1984gx} 
  S.~Dawson,
  Nucl.\ Phys.\ B {\bf 249}, 42 (1985).

\bibitem{Boos:1997gw} 
  E.~Boos, H.~J.~He, W.~Kilian, A.~Pukhov, C.~P.~Yuan and P.~M.~Zerwas,
  Phys.\ Rev.\ D {\bf 57}, 1553 (1998)
    [hep-ph/9708310].


\bibitem{Kleiss:1986xp} 
  R.~Kleiss and W.~J.~Stirling,
  Phys.\ Lett.\ B {\bf 182}, 75 (1986).

\bibitem{Kunszt:1987tk} 
  Z.~Kunszt and D.~E.~Soper,
  Nucl.\ Phys.\ B {\bf 296}, 253 (1988).

\bibitem{Accomando:2006mc} 
  E.~Accomando, A.~Ballestrero, A.~Belhouari and E.~Maina,
  Phys.\ Rev.\ D {\bf 74}, 073010 (2006)
  [hep-ph/0608019].

\bibitem{Alboteanu:2008my} 
  A.~Alboteanu, W.~Kilian and J.~Reuter,
  JHEP {\bf 0811}, 010 (2008)
    [arXiv:0806.4145 [hep-ph]].

\bibitem{Borel:2012by} 
  P.~Borel, R.~Franceschini, R.~Rattazzi and A.~Wulzer,
  JHEP {\bf 1206}, 122 (2012)
    [arXiv:1202.1904 [hep-ph]].


\bibitem{Lindfors:1985yp} 
  J.~Lindfors,
  Z.\ Phys.\ C {\bf 28}, 427 (1985).

\bibitem{Johnson:1987tj} 
  P.~W.~Johnson, F.~I.~Olness and W.~K.~Tung,
  Phys.\ Rev.\ D {\bf 36}, 291 (1987).

\bibitem{Kuss:1995yv} 
  I.~Kuss and H.~Spiesberger,
  Phys.\ Rev.\ D {\bf 53}, 6078 (1996)
    [hep-ph/9507204].

\bibitem{Kuss:1996ww} 
  I.~Kuss,
  Phys.\ Rev.\ D {\bf 55}, 7165 (1997)
  [hep-ph/9608453].

\bibitem{Bernreuther:2015llj} 
  W.~Bernreuther and L.~Chen,
  Phys.\ Rev.\ D {\bf 93}, 053018 (2016)
   [arXiv:1511.07706 [hep-ph]].


\bibitem{Accomando:2006hq}
  E.~Accomando, A.~Denner and S.~Pozzorini,
  JHEP {\bf 0703} (2007) 078
  [hep-ph/0611289].

\bibitem{Bouayed:2007rt} 
  N.~Bouayed and F.~Boudjema,
  Phys.\ Rev.\ D {\bf 77}, 013004 (2008)
   [arXiv:0709.4388 [hep-ph]].

\bibitem{Frixione:1993yw} 
  S.~Frixione, M.~L.~Mangano, P.~Nason and G.~Ridolfi,
  Phys.\ Lett.\ B {\bf 319}, 339 (1993)
   [hep-ph/9310350].

\bibitem{okun}
See, e.g., L.~B.~Okun, Leptons and Quarks, North Holland, 1984.


\bibitem{Alikhanov:2008eu} 
  I.~Alikhanov,
  Eur.\ Phys.\ J.\ C {\bf 56}, 479 (2008)
  [arXiv:0803.3707 [hep-ph]].

\bibitem{Alikhanov:2008ca} 
  I.~Alikhanov,
  Eur.\ Phys.\ J.\ C {\bf 65}, 269 (2010)
  [arXiv:0812.0937 [hep-ph]].


\bibitem{Chen:1974wv} 
  M.~S.~Chen and P.~M.~Zerwas,
  Phys.\ Rev.\ D {\bf 11}, 58 (1975).


\bibitem{Flambaum:1974wp} 
  V.~V.~Flambaum, I.~B.~Khriplovich and O.~P.~Sushkov,
  Sov.\ J.\ Nucl.\ Phys.\  {\bf 20}, 537 (1975)
  [Yad.\ Fiz.\  {\bf 20}, 1016 (1974)].

\bibitem{Berends:1986yy} 
  F.~A.~Berends, G.~J.~H.~Burgers and W.~L.~van Neerven,
  Phys.\ Lett.\ B {\bf 177}, 191 (1986).

\bibitem{Mery:1987et} 
  P.~Mery, M.~Perrottet and F.~M.~Renard,
  Z.\ Phys.\ C {\bf 38}, 579 (1988).

\bibitem{Baier:1969kaa} 
  V.~N.~Baier and V.~S.~Fadin,
  Phys.\ Lett.\ B {\bf 27}, 223 (1968).


\bibitem{Brown:1978mq} 
  R.~W.~Brown and K.~O.~Mikaelian,
  Phys.\ Rev.\ D {\bf 19}, 922 (1979).

\bibitem{Renton:1990td} 
  See, e.g., P.~Renton,
  Electroweak Interactions: An Introduction to the Physics of Quarks and Leptons,
  Cambridge University Press, 1990.

\bibitem{Alikhanov:2018jey} 
  I.~A.~Alikhanov,
  Phys.\ Part.\ Nucl.\  {\bf 49}, 670 (2018).
	
\bibitem{Bednyakov:2015uoa} 
  V.~A.~Bednyakov,
  Phys.\ Part.\ Nucl.\  {\bf 47}, 711 (2016)
   [arXiv:1505.04380 [hep-ph]].


\end{thebibliography}
\end{document}